\documentclass{aa}

\usepackage{natbib}
\bibpunct{(}{)}{;}{a}{}{,} 

\usepackage{graphicx}

\usepackage{amssymb,amsmath,bm}
\usepackage{delarray}
\usepackage{mathrsfs}
\usepackage{xcolor}
\usepackage{soul}
\usepackage[varg]{txfonts}
\usepackage{hyperref}
\hypersetup{
  colorlinks   = true, 
  urlcolor     = blue, 
  linkcolor    = blue, 
  citecolor    = blue  
}

\begin{document} 

\title{Diagnostic potential of wavelength-integrated scattering polarisation signals of the
solar He~{\sc ii} Ly-$\alpha$ line}

\author{Fabio Riva\inst{1,2}
       \and
       Gioele Janett\inst{1,2}
       \and
       Luca Belluzzi\inst{1,2,3}
       }

\institute{
        Istituto ricerche solari Aldo e Cele Daccò (IRSOL), Faculty of Informatics, Università della Svizzera italiana (USI), CH-6605 Locarno, Switzerland 
        \and
        Euler Institute, Universit\`a della Svizzera italiana (USI), CH-6900 Lugano, Switzerland
        \\ \email{fabio.riva@irsol.usi.ch}
        \and
        Leibniz-Institut f\"ur Sonnenphysik (KIS), D-79104 Freiburg i.~Br., Germany\\          
        }

\date{Received xxx; accepted yyy}
 
\abstract
{}
%
{
The main goal of this work is to study the potential of He~{\sc ii}~Ly-$\alpha$ wavelength-integrated scattering polarisation for probing the magnetism 
of the solar upper chromosphere.
Meanwhile, the suitability of different modelling approximations 
is investigated.
}
%
{
Radiative transfer calculations are performed in semi-empirical 1D solar atmospheres, out of local thermodynamic equilibrium, 
considering a two-term atomic model and accounting for the Hanle, Zeeman, and magneto-optical effects.
The problem is suitably linearised and discretised, and the resulting numerical system is solved with a matrix-free iterative method.
The results obtained modelling scattering processes with three different descriptions, namely in the limit of complete frequency redistribution (CRD), and accounting for partial frequency redistribution (PRD) effects under the angle-averaged (AA) approximation and in the general angle-dependent (AD) formulation, are compared.
}
%
{
In the line-core, the synthetic Stokes profiles resulting from CRD, PRD--AA, and PRD--AD calculations show a very good agreement.
On the other hand, relevant differences are observed in $Q/I$ outside the line-core region.
Besides, the precise structure of the atmospheric model does not noticeably affect the line-core profiles, but it strongly 
impacts the $Q/I$ signals outside the line-core.
As most of the He~{\sc ii}~Ly-$\alpha$ photons originate in the core region, it turns out that 
wavelength-integrated linear polarisation signals are almost insensitive to both the scattering 
description and the atmospheric model.
Appreciable wavelength-integrated $\overline{U}/\overline{I}$ 
signals, showing observable 
sensitivity to horizontal magnetic fields in the range 0-1000~G, are also found, particularly near the limb.
It turns out that, while the integration time required to detect magnetic fields in the quiet chromosphere with this line is too long for sounding rocket missions, magnetic fields corresponding to typical 
plage areas would produce detectable signals, especially near the limb.
}
%
{
These results, to be confirmed by 3D calculations that will include the impact of horizontal inhomogeneities and bulk velocity gradients, show that filter-polarimetry 
in the He~{\sc ii} Ly-$\alpha$ line has a promising potential for chromospheric magnetic field diagnostics.
In near-limb plage regions, 
this could already be assessed through sounding rocket experiments.

}



\maketitle

\section{Introduction}\label{sec:introduction}

The dynamics of the outer layers of the solar atmosphere is largely governed by the presence of magnetic fields that, however, 
are notoriously hard to measure.
The most promising state-of-the-art remote sensing approach to overcome this difficulty and 
explore the magnetism of the solar upper 
chromosphere and transition region
is the interpretation of the
polarisation signals produced by both the Zeeman effect and scattering processes in strong UV resonance lines, such as Mg~{\sc ii} $h\&k$, 
H~{\sc i} Ly-$\alpha$, or He~{\sc ii} Ly-$\alpha$ 
\citep[e.g.][]{trujillo_bueno2022}.
Unfortunately, the Zeeman effect
is not very effective at short wavelengths and in hot plasmas. Thus, the signals
produced by the weak magnetic fields of the quiet solar chromosphere through this mechanism can hardly be detected in far and extreme UV lines.
On the other hand, all the aforementioned resonance lines show measurable linear scattering polarization signals, with an interesting 
magnetic sensitivity via the combined action of the Hanle and magneto-optical (MO) effects.

Thanks to its sensitivity to chromospheric magnetic fields, the H~{\sc i} Ly-$\alpha$ line
at 1216~\AA~attracted increasing attention during the last few years.
Modelling scattering processes in the limit of
complete frequency redistribution (CRD)
and neglecting $J$-state interference,
\citet{trujillo_bueno2011} showed that
the line-core scattering polarisation signal is sensitive, via the Hanle effect, to the presence of magnetic fields of strengths between 
10~G and 100~G in the upper chromosphere.
\citet{stepan2012} and \citet{stepan2015} extended the investigation of the Hanle sensitivity of the H~{\sc i} Ly-$\alpha$ line
by considering realistic 2D and 3D atmospheric models resulting from a state-of-the-art radiation magneto-hydrodynamic simulation.
A very interesting outcome of these investigations is that the centre-to-limb variation~(CLV) of the spatially averaged line-core $Q/I$ signal
is qualitatively similar to that found considering semi-empirical 1D atmospheric models.
Moreover, \citet{belluzzi2012b} showed that the joint action of partial frequency redistribution (PRD) and $J$-state interference effects produces conspicuous 
wing scattering polarisation signals, which are very sensitive to the chromospheric thermal structure.
These results motivated the development of the Chromospheric Lyman-Alpha Spectro-Polarimeter~(CLASP)
sounding rocket experiment 
\citep{kobayashi2012,kano2012},
which 
paved the way to the 
observation of the intensity and linear
polarisation of the solar H~{\sc i} Ly-$\alpha$ line.
The CLASP data showed a remarkably good qualitative agreement with the theoretical predictions \citep{kano2017} and have been exploited to get
new constraints on the magnetization and the degree of
corrugation of the chromosphere-corona transition region \citep{trujillo_bueno2018}.
More recently,
\citet{alsinaballester2018} and~\citet{alsina_ballester2023} proved that
the wavelength-integrated linear polarisation signals
of this line are strongly
sensitive to magnetic fields of strengths of
about 50\,G in the middle–upper chromosphere via MO
effects.

By contrast, a reduced number of investigations has been dedicated to the scattering polarisation signals of the 
He~{\sc ii} Ly-$\alpha$ line at 304~\AA~and
especially to its magnetic sensitivity.
This for two reasons.
First, the He~{\sc ii} Ly-$\alpha$ line is much less sensitive to the Hanle effect than the H~{\sc i} Ly-$\alpha$ line.
Indeed, having a much larger Einstein coefficient for spontaneous emission, the Hanle critical field of the He~{\sc ii} Ly-$\alpha$ line is $B_\text{H} \approx 850$\,G, 
while the H~{\sc i} Ly-$\alpha$ line has $B_\text{H} \approx 53$\,G.
By applying the CRD limit, \citet{trujillo_bueno2012} confirmed that the He~{\sc ii} Ly-$\alpha$ line-core $Q/I$ signal is nearly immune to magnetic fields weaker than 100\,G,
thus proposing to use it as a reference signal for a differential Hanle effect technique.
Second, although the He~{\sc ii} Ly-$\alpha$ line shows large fractional linear polarisation  signals,
the required exposure time to reach any given polarimetric sensitivity is significantly larger than for the 
H~{\sc i} Ly-$\alpha$ line \citep[e.g.][]{trujillo_bueno2022}.

The scattering polarisation signals of the He~{\sc ii} Ly-$\alpha$ line in unmagnetised atmospheres
were further explored
by \citet{belluzzi2012b},
who showed that the joint action of PRD
and $J$-state interference effects produces complex scattering polarisation $Q/I$ profiles
with sharp peaks of large amplitude outside the line-core region.
Although these peaks are largely insensitive to the Hanle effect,
in principle they could be
impacted by magnetic fields through the action of MO 
effects 
\citep[e.g.][]{alsinaballester2016,delpinoaleman2016}.
This potential, yet unexplored, magnetic sensitivity of the He~{\sc ii} Ly-$\alpha$ line motivates the work in the present paper.
Moreover, we note that, for computational simplicity,
the PRD calculations of 
\citet{belluzzi2012b} 
were performed within the angle-averaged (AA) approximation
\citep[e.g.][]{mihalas1978,leenaarts2012,sampoorna2017,alsinaballester2017}. 
The suitability of such an approximation in the modelling of He~{\sc ii} Ly-$\alpha$ line polarisation signals has yet to be proven.

Motivated by the studies of \citet{trujillo_bueno2012} and \citet{belluzzi2012b}, the main goal of this work is to investigate the potential of He~{\sc ii} Ly-$\alpha$ wavelength-integrated scattering polarisation
for probing the magnetism of the upper solar chromosphere.
Concurrently, we aim to assess the suitability 
of the CRD and PRD--AA approximations to model the polarization profiles and the wavelength-integrated polarization signals of this line. This is done through a comparison with general angle-dependent~(AD) PRD calculations, which are more accurate but, at 
the same time, also imply a significantly higher computational cost. 

The present article is organised as follows. 
Section~\ref{sec:problem} exposes the considered non-equilibrium transfer problem for polarised radiation, as well as its linearization, discretization, 
and algebraic formulation.
Section~\ref{sec:setting} presents the adopted numerical solution strategy and the numerical set-up.
In Sect.~\ref{sec:numerical_results}, we report and analyse the synthetic emergent Stokes profiles of the He~{\sc ii} Ly-$\alpha$ line,
comparing CRD, PRD--AA, and PRD--AD calculations and investigating the impact that the magnetic field and the atmospheric parameters have on the numerical results.
Finally, Sect.~\ref{sec:conclusions} discusses the main results and their implications, and provides some final considerations.

\section{Formulation of the problem}
\label{sec:problem}

Aiming to model the scattering polarisation signal of the He~{\sc ii} Ly-$\alpha$ line at 304~\AA,
we solved the non-equilibrium transfer problem for polarised radiation in 1D models of the solar atmosphere.
To correctly model this signal, a two-term atomic model, which includes quantum interference between different fine-structure~(FS) levels 
of the same term \citep[$J$-state interference, see][]{landi_deglinnocenti+landolfi2004} has to be considered \citep{belluzzi2012b}.
Scattering processes are modelled considering three different descriptions 
of increasing computational complexity: 
(i) the CRD limit; and including PRD effects
(ii) within the AA approximation; or (iii) in their general AD formulation.
We also consider the impact of magnetic fields of arbitrary strength and orientation, accounting for the Hanle, 
Zeeman, and MO
effects.
The theoretical framework is presented in the following.

\subsection{Two-term atomic model}
\label{sec:atomic_model}

The He~{\sc ii} Ly-$\alpha$ line results from two distinct FS transitions between the ground level of ionized helium $1s \, ^2\mathrm{S}_{1/2}$ and the 
excited levels $2p \, ^2\mathrm{P}_{1/2}^\mathrm{\, o}$ and $2p \, ^2\mathrm{P}_{3/2}^\mathrm{\, o}$.
These two FS components are separated by 5\,m{\AA} only, and are thus completely blended.
To take into account both these transitions, as well as the $J$-state interference between the two upper levels, we consider a two-term 
$^2\mathrm{S}$ -- $^2\mathrm{P}^\mathrm{\, o}$ atomic model.
Noticing that the lower level has $J=1/2$ and therefore cannot carry atomic alignment \citep[see][]{landi_deglinnocenti+landolfi2004}, and that it has a very long lifetime, 
we can assume that it is unpolarized and infinitely sharp.
Neglecting stimulated emission, which is generally a good assumption in the solar atmosphere, the statistical equilibrium equations for a two-term atom with 
unpolarized and infinitely-sharp lower term have an analytic solution.
Thus, the scattering contribution to the line emissivity can be expressed through the redistribution matrix formalism, which is a very convienient formulation for modelling PRD 
effects. 
In this work, we used the redistribution matrix for the aforementioned atomic system as derived by \citet{bommier2017} (see Sect.~\ref{sec:scattering_integral}).
For the energies of the upper levels $^2\mathrm{P}_{1/2}^\mathrm{\, o}$ and $^2\mathrm{P}_{3/2}^\mathrm{\, o}$, relative to the ground state, we considered $E_u=329179.29\ \text{cm}^{-1}$ and 
$E_u=329185.15\ \text{cm}^{-1}$, respectively, while for the Einstein coefficient for spontaneous emission from upper to lower term, we used 
$A_{ul}=1.0029\cdot10^{10}\ \text{s}^{-1}$.
The splitting of the magnetic sublevels in the presence of a magnetic field was calculated considering the Paschen-Back effect.
The theoretical values of the Land\'e factors were used.
We finally note that only the transition from the upper level $^2\mathrm{P}_{3/2}^\mathrm{\, o}$ contributes to the emergent linear scattering polarization \citep{trujillo_bueno2011}.

\subsection{Semi-empirical 1D atmospheric models}
\label{sec:atmospheric_model}

The calculations presented in the following were 
obtained with the plane-parallel semi-empirical 1D models of
\citet[][hereafter FAL models]{fontenla1993},
making use of the microturbulent velocities as determined
in \citet{fontenla1991}.
In particular, we considered three quiet Sun atmospheric models, representing a faint inter-network region (FAL-A),
a bright network region (FAL-F), and an average region (FAL-C), as well as a model that mimics a plage area (FAL-P), that is a region that appears particularly bright in the chromospheric hydrogen H-$\alpha$ line, typically found next to active 
solar regions.
Despite their simplicity, these atmospheric models are realistic enough to assess the potential of He~{\sc ii} Ly-$\alpha$ wavelength-integrated 
scattering polarisation for investigating chromospheric magnetic fields.

\subsection{Radiative transfer equation}
\label{sec:RT}

The intensity and polarisation of a beam of radiation are fully described by the Stokes vector $\bm{I}\in\mathbb R^4$.
Considering a Cartesian reference system with the $z$-axis (quantization axis for total angular momentum)
directed along the vertical,
the transfer of partially polarised radiation is described
by the following differential equation
\begin{equation}\label{eq:RT}
        \cos(\theta)\frac{\rm d}{{\rm d} z}\bm{I}(z,\mathbf{\Omega},\nu) = 
	-
	K(z,\mathbf{\Omega},\nu) 
	\bm{I}(z,\mathbf{\Omega},\nu) + 
	\bm{\varepsilon}(z,\mathbf{\Omega},\nu),
\end{equation}
where the vector $\vec{\Omega}=(\theta,\chi)\in [0,\pi]\times[0,2\pi)$
and the scalar $\nu$ specify the direction and the frequency, respectively, of the considered radiation beam.
The inclination $\theta$ is measured with respect to the 
$z$-axis and corresponds to the heliocentric angle (generally specified by $\mu=\cos{\theta} \in [-1,1]$) of the observed point.
The azimuth $\chi$ is measured from the $x$-axis counter-clockwise for an observer at positive $z$.
The propagation matrix $K\in\mathbb R^{4\times4}$ describes how the medium differentially absorbs the polarised radiation (dichroism)
and how it couples the different Stokes parameters (MO effects).
The emission vector $\bm{\varepsilon}\in\mathbb R^4$ describes the polarised radiation emitted by the solar plasma.

In the following, we consider the contributions to $K$ and $\vec{\varepsilon}$ due to both line and continuum processes.
These contributions, labelled with the superscripts $\ell$ and $c$, respectively, simply sum up.
Moreover, both $\vec{\varepsilon}^\ell$ and $\vec{\varepsilon}^c$ can in turn be written as the sum of two terms, describing the contributions from 
scattering (label ``$\text{sc}$'') and thermal (label ``$\text{th}$'') processes, respectively.
Hence, we have
\begin{equation*}
        K(z,\mathbf{\Omega},\nu) =
        K^c(\vec{r},\mathbf{\Omega},\nu)
        + K^{\ell}(z,\mathbf{\Omega},\nu),
\end{equation*}
and
\begin{align}\label{eq:eps_sc_th}
	\bm{\varepsilon}(z,\mathbf{\Omega},\nu) = & 
	\bm{\varepsilon}^{c,\text{th}}(z,\mathbf{\Omega},\nu) + 
        \bm{\varepsilon}^{c,\text{sc}}(z,\mathbf{\Omega},\nu)\\
        + &
	\bm{\varepsilon}^{\ell,\text{th}}(z,\mathbf{\Omega},\nu)+
	\bm{\varepsilon}^{\ell,\text{sc}}(z,\mathbf{\Omega},\nu).\nonumber
\end{align}
The explicit expressions for $K^c$ and $K^{\ell}$
are given by \cite{landi_deglinnocenti+landolfi2004},
while we refer to the work of \citet{alsinaballester2022}
for $\bm{\varepsilon}^{c,\text{th}}$,
$\bm{\varepsilon}^{c,\text{sc}}$,
and $\bm{\varepsilon}^{\ell,\text{th}}$.
The line scattering contribution to emissivity, $\bm{\varepsilon}^{\ell,\text{sc}}$, is described in the following section.

\subsection{Line scattering emissivity}
\label{sec:scattering_integral}

Using the redistribution matrix formalism and following the convention that primed and unprimed quantities refer to the incident and scattered radiation, respectively,
the line scattering term can be written through the scattering integral
\begin{equation}\label{eq:eps_sc_l}
	\bm{\varepsilon}^{\ell,\text{sc}}(z,\bm{\Omega},\nu) \!=\! 
	k_M(z)\!\! \int\!\! {\rm d} \nu'\! 
	\oint\! \frac{{\rm d} \bm{\Omega}'}{4 \pi}
	R(z,\bm{\Omega}',\bm{\Omega},\nu',\nu) 
	\bm{I}(z,\bm{\Omega}',\nu'),
\end{equation}
where the factor $k_M$ is the wavelength-integrated absorption coefficient \citep[e.g.][]{alsinaballester2022} and $R \in \mathbb{R}^{4 \times 4}$ is the redistribution matrix
for the considered two-term atomic model \citep[see][]{bommier2017}. We recall that $R$ encodes the statistical equilibrium (SE) equations, which determine the state of the atomic system.

The redistribution matrix can be expressed as the sum of two terms, 
$R=R^{\scriptscriptstyle \rm II}+R^{\scriptscriptstyle \rm III}$, which describe scattering processes that are coherent ($\rm II$) and completely incoherent ($\rm III$) in the atomic 
reference frame. 
As far as $R^{\scriptscriptstyle \rm III}$ is concerned, we considered its simplified expression under the assumption of CRD in the observer's frame 
\citep[see][for a discussion about its suitability]{bommier1997b,sampoorna2017,alsinaballester2017,riva2023}.
For conciseness, we do not provide the explicit expressions of $R^{\scriptscriptstyle \rm II}$ and $R^{\scriptscriptstyle \rm III}$. Instead, we refer to the work of \citet{alsinaballester2022} for the analytical form of $R^{\scriptscriptstyle \rm II}$ in the observer's frame, both in its 
general AD formulation and under the AA approximation, as well as for the expression of $R^{\scriptscriptstyle \rm III}$, under the aforementioned assumption of CRD in the observer's frame.
Moreover, we note that the limit of CRD is obtained by artificially increasing the rate of elastic collisions in the branching ratios of $R^{\scriptscriptstyle \rm II}$ and 
$R^{\scriptscriptstyle \rm III}$ so that the contribution of the former vanishes \citep[see][]{alsinaballester2022}. 
In this respect, we recall that the considered redistribution matrix does not account for the depolarizing effect of elastic collisions with neutral perturbers, 
because the SE equations for a two-term atom in the presence of magnetic fields cannot be solved analytically when such effect is included 
\citep[see][for more details]{bommier2017}.
Noticing that \citet{alsinaballester2021} verified that this effect can be safely ignored for the Na~{\sc i} D lines, we can argue that the same conclusion also 
holds for the He~{\sc ii} Ly-$\alpha$ line, which forms significantly higher in the solar atmosphere.

\subsection{Linearization, discretization, and algebraic formulation}

The radiative transfer (RT) problem consists in finding a self-consistent solution of the RT equation 
for the radiation field $\bm{I}$ and of the SE equations for the atomic system, whose state determines $K$ and $\bm{\varepsilon}$.
The problem is in general integro-differential, nonlocal, and nonlinear.
In our formulation, based on the redistribution matrix formalism, the nonlinearity
lies in the dependence of both $K$ and $\bm{\varepsilon}$ on the coefficient $k_M$. 
This coefficient is proportional to the population of the lower term, which in turn nonlinearly depends on $\bm{I}$ through 
the SE equations
\citep[see e.g.][]{landi_deglinnocenti+landolfi2004}.
However, the RT problem can
be suitably linearised
by fixing a priori the coefficient $k_M$, which 
corresponds to assuming a fixed
lower term population
\citep[see e.g.][]{belluzzi2014,sampoorna2017,alsinaballester2017,janett2021b,benedusi2021,benedusi2022}.
This assumption makes the propagation matrix $K$ independent of $\bm{I}$, whereas $\bm{\varepsilon}$ depends on $\bm{I}$ only linearly.
In this case, the whole problem consists in finding a self-consistent solution of Eqs.~(\ref{eq:RT})-(\ref{eq:eps_sc_l}) and it is
linear in $\bm{I}$.

To numerically solve the resulting linearised RT problem, we first need to discretise the continuous variables 
$z$, $\mathbf{\Omega}$, and $\nu$.
The spatial discretization is 
done by using
$N_r$ unevenly spaced grid nodes.
The angular grid, which samples the unit sphere
with $N_\Omega$ angular nodes,
usually depends on the quadrature chosen to evaluate the angular integral in Eq.~\eqref{eq:eps_sc_l}
(see Sect.~\ref{sec:quadratures}).
Finally, we consider a finite spectral interval $[\nu_{\min},\nu_{\max}]$ for $\nu$, discretised with $N_\nu$ unevenly spaced nodes. 
More details on the numerical parameters used for the present work are given in Sect.~\ref{sec:physical_parameters}.

The discretised radiation field and emissivity vectors are thus represented by $\mathbf{I}\in\mathbb R^N$ and $\bm{\varepsilon}\in\mathbb R^N$, 
with $N=4N_rN_\nu N_\Omega$ the total number of degrees of freedom. 
Hence, we can express the transfer equation, Eq.~\eqref{eq:RT}, and the
calculation of the emission coefficient, Eq.~\eqref{eq:eps_sc_th}, as
\begin{align}
\mathbf{I}&=\Lambda\pmb{\varepsilon}+\mathbf{t},
\label{eq:matrix_form_1}\\
\pmb{\varepsilon}&=\Sigma\mathbf{I}+\pmb{\varepsilon}^{\text{th}},
\label{eq:matrix_form_2}
\end{align}
respectively. 
Here, the transfer operator
$\Lambda\in\mathbb R^{N\times N}$ and the scattering operator $\Sigma\in\mathbb R^{N\times N}$ are linear, the vector $\mathbf{t}\in\mathbb R^N$ represents the radiation transmitted from the boundaries,
while $\pmb{\varepsilon}^{\text{th}}\in\mathbb R^N$
represents the thermal contributions
to the emissivity.
The evaluation of the term $\Lambda\pmb{\varepsilon}+\mathbf{t}$ is commonly known as formal solution
and consists in the numerical solution of Eq.~\eqref{eq:RT} for $\mathbf{I}$, provided $\pmb{\varepsilon}$, $K$, and appropriate boundary conditions.
The term $\Sigma\mathbf{I}$ corresponds to the numerical evaluation of the scattering contribution to the emission vector, given $\mathbf{I}$. 
This requires approximating the integral in Eq.~\eqref{eq:eps_sc_l} via suitable quadrature rules (see Sect.~\ref{sec:quadratures}).
Equations~\eqref{eq:matrix_form_1} and~\eqref{eq:matrix_form_2} can then be combined into a matrix equation of the form
$A\mathbf{x}=\mathbf{b}$, namely, 
\begin{equation}\label{eq:linear_system}
(Id-\Lambda\Sigma)\mathbf{I}=\Lambda\bm{\varepsilon}^{\text{th}}+\mathbf{t},
\end{equation}
being $Id\in\mathbb R^{N\times N}$ the identity matrix \citep[see e.g.][]{janett2021b,benedusi2022}.

\section{Methods and setting}
\label{sec:setting}

We now expose the numerical methods employed to
solve Eq.~\eqref{eq:linear_system} and the 
physical and numerical parameters used in the calculations.

\subsection{Angular and spectral quadratures}
\label{sec:quadratures}

The numerical integration of Eq.~\eqref{eq:eps_sc_l} requires a proper choice of quadrature rules.
For the angular integral,
we resorted on a tensor product
of a trapezoidal quadrature in the azimuthal interval
$[0,2\pi)$ for $\chi$
and two Gauss-Legendre quadratures
in each inclination subinterval $(-1,0)$ and $(0,1)$
for $\mu=\cos(\theta)$.
For the integral in $\nu'$, we selected a set of spectral points
distributed ad hoc to approximately follow the shape of the redistribution matrix $R$. More precisely, we used a Gauss-Hermite quadrature rule in spectral regions where $R$ is similar to a Gaussian function, while we used a Gauss-Legendre quadrature rule in the rest of the parameter domain.

\subsection{Formal solution}
\label{sec:formal_solution}

For a better agreement with the results by \citet{belluzzi2012b},
we applied the DELOPAR formal solver 
to evaluate the $\Lambda$ operator in Eq.~\eqref{eq:linear_system} \citep[for details, see][]{trujillo_bueno2003,janett2017formal,janett2018formal,janett2018formal2},
using a trapezoidal quadrature for the conversion to optical depth.
We note that the use of the $L$-stable DELO-linear method
only affects the amplitude of the emergent Stokes profiles,
but leads to the same general conclusions.

\subsection{Iterative solver}
\label{sec:iterative_method}

The whole transfer problem has been recast
in the linear system given by Eq.~\eqref{eq:linear_system},
for which we have to find a solution.
Given the size of the system, with $N\sim10^6$ (see Sect.~\ref{sec:physical_parameters}), the use of direct matrix inversion solvers
is definitely unfeasible. Thus,
the application of an iterative method
was preferred.
Moreover, it is
also too expensive to explicitly assemble
the dense operator $Id-\Lambda\Sigma$
in~Eq.~\eqref{eq:linear_system}.
Therefore, the action of the operator $Id-\Lambda\Sigma$ was encoded in a routine
and an iterative method was applied
in a matrix-free context. 

In this work, we applied a Krylov solver, namely the generalised  minimal residual (GMRES) method,
which proved to be highly effective in terms of convergence rate and time-to-solution
for this kind of problem
\citep{benedusi2021,benedusi2022,benedusi2023}. For the PRD calculations, we additionally exploited
the innovative approach to designing
efficient physics-based preconditioners and initial guesses
extensively described in~\citet{janett2023} and we monitored the convergence of
the GMRES iterative method by considering
the preconditioned relative residual
\begin{equation*}
{\rm res} = \frac{\|P^{-1}
(\Lambda\bm{\varepsilon}^{\text{th}}+\mathbf{t}-
(Id-\Lambda\Sigma)\mathbf I)\|_2}{\|P^{-1}(\Lambda\bm{\varepsilon}^{\text{th}}+\mathbf{t})\|_2},
\end{equation*}
with $P$ the preconditioner, taking as a stopping criterion ${\rm res}<{\rm tol}=10^{-9}$.
In the calculations presented in this paper,
both CRD- and PRD--AA-based preconditioners
have a noticeable impact,
ensuring a solution of
the problem in 4-8 iterations.
Moreover, the inclusion of polarisation
in the preconditioner has small impact
(with respect to the unpolarised version)
and it is thus neglected.

\subsection{Physical and numerical parameters}
\label{sec:physical_parameters}

The lower term population was provided by
the FAL atmospheric models \citep[][]{fontenla1993}.
The RH code by \citet{uitenbroek2001} was used
to calculate
the continuum total opacity,
scattering opacity, and thermal emissivity, the damping parameter, as well as the collisional rates, except for the one for
inelastic de-exciting collisions, which was evaluated as $C_{ul}=N_eQ_{ul}$, with $N_e$ the electron density and 
$$Q_{ul}=8.63\cdot10^{-6}\cdot\upsilon/(g_u\sqrt{T}),$$
where $g_u=6$ is the degeneracy of the upper term, $T$ is the local temperature, and $\upsilon$ (which depends on $T$) 
is taken from \citet{janev1987}.
For the formal solution, we assumed the following boundary conditions
\begin{align*}
I(z_{\min},\theta,\chi,\nu) &= B_P,\qquad
&&\text{ for }\;\theta\in [0, \pi/2), \forall \chi, \forall\nu,\\
I(z_{\max},\theta,\chi,\nu) &= 0,\qquad
&&\text{ for }\;\theta\in (\pi/2, \pi], \forall \chi, \forall\nu,
\end{align*}
where $B_P$ is the line-centre frequency Planck function at the local temperature.

The wavelength interval
$[\lambda_{\min},\lambda_{\max}]=[303.01\text{\,\AA},304.55\text{\,\AA}]$
was discretised with ${N_\nu=141}$ frequency nodes,
approximately uniformly spaced in the line core and logarithmically distributed in the wings. 
The spatial grid was provided by the considered
semi-empirical 1D atmospheric models. However, since the He~{\sc ii} Ly-$\alpha$ line forms in a very narrow region in the outermost layers of the solar chromosphere~\citep[e.g.][]{trujillo_bueno2012,belluzzi2012b}, we decided to cut out the layers below $z_\text{min}=1375$, 1475, 1475, and 1495 km for the FAL-A, FAL-C, FAL-F, and FAL-P models, respectively, as their contribution to the emergent radiation field is negligible.\footnote{We repeated part of the calculations with the full FAL models, finding undetectable differences.} Conversely, we doubled the spatial resolution in the uppermost layers. This resulted in 61-76 unevenly distributed spatial nodes, depending on the considered atmospheric model.
For the angular discretization, we used
two 6-nodes Gauss-Legendre grids
for the inclination and a 9-nodes grid
for the azimuth.


\section{Results}
\label{sec:numerical_results}

\begin{figure*}[ht!]
    \centering
    \includegraphics[width=0.95\textwidth]{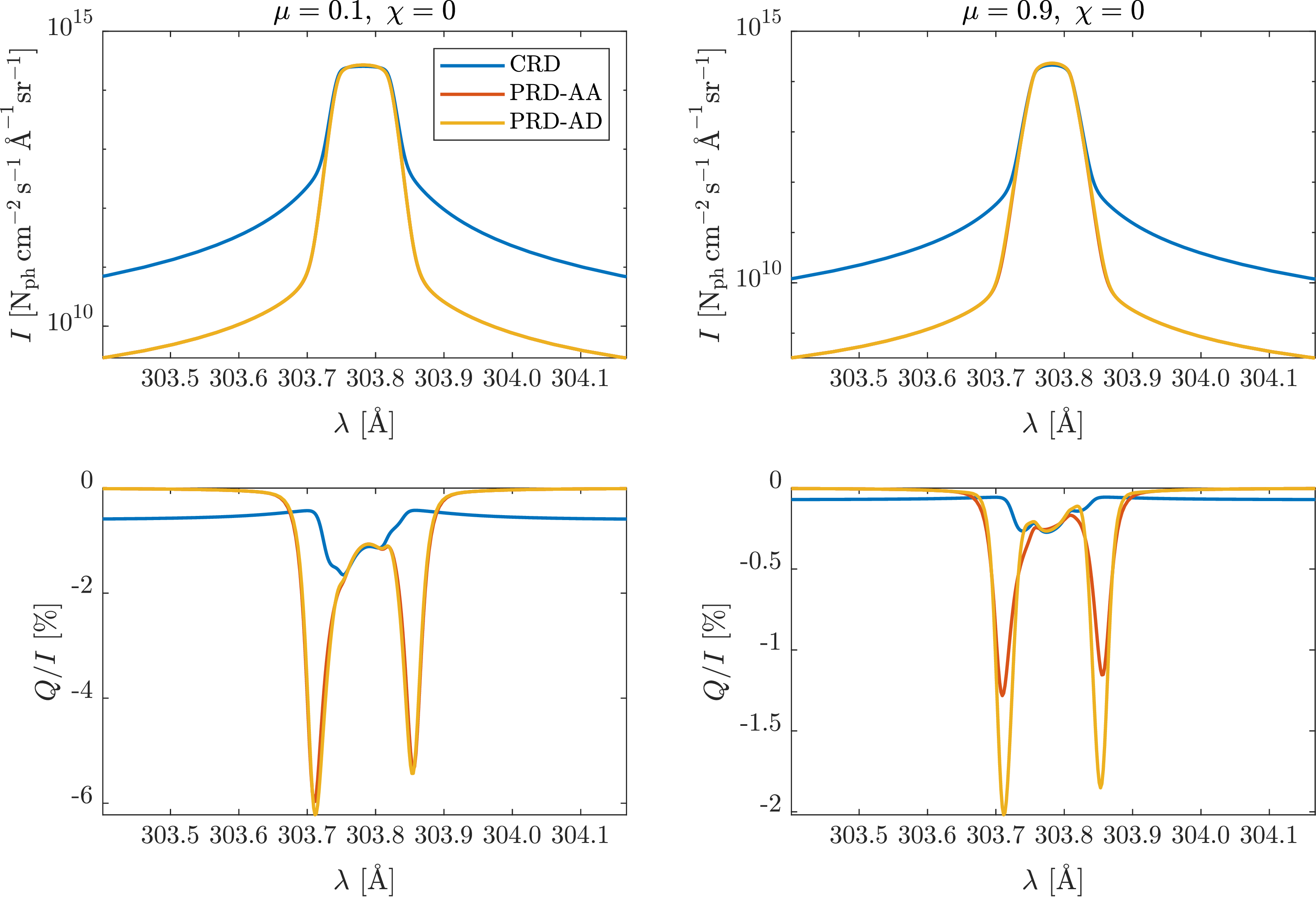}
    \caption{Stokes $I$ (\emph{top panels}) and $Q/I$ (\emph{bottom panels}) profiles for the He~{\sc ii} Ly-$\alpha$ line as a function of wavelength. The results are obtained with the unmagnetised FAL-C atmospheric model and for two LOSs, near the limb ($\mu=0.1$, \emph{left panels}) and near the disk centre ($\mu=0.9$, \emph{right panels}), respectively. Blue, red, and yellow lines correspond to CRD, PRD--AA, and PRD--AD calculations,
    respectively. The intensity is in units of number of photons per unit of surface, time, wavelength, and solid angle. The intensity profiles for PRD--AA and PRD--AD calculations overlap.}
    \label{fig:I_QI_unmagnetic}
\end{figure*}

\begin{figure*}[ht!]
    \centering
    \includegraphics[width=0.95\textwidth]{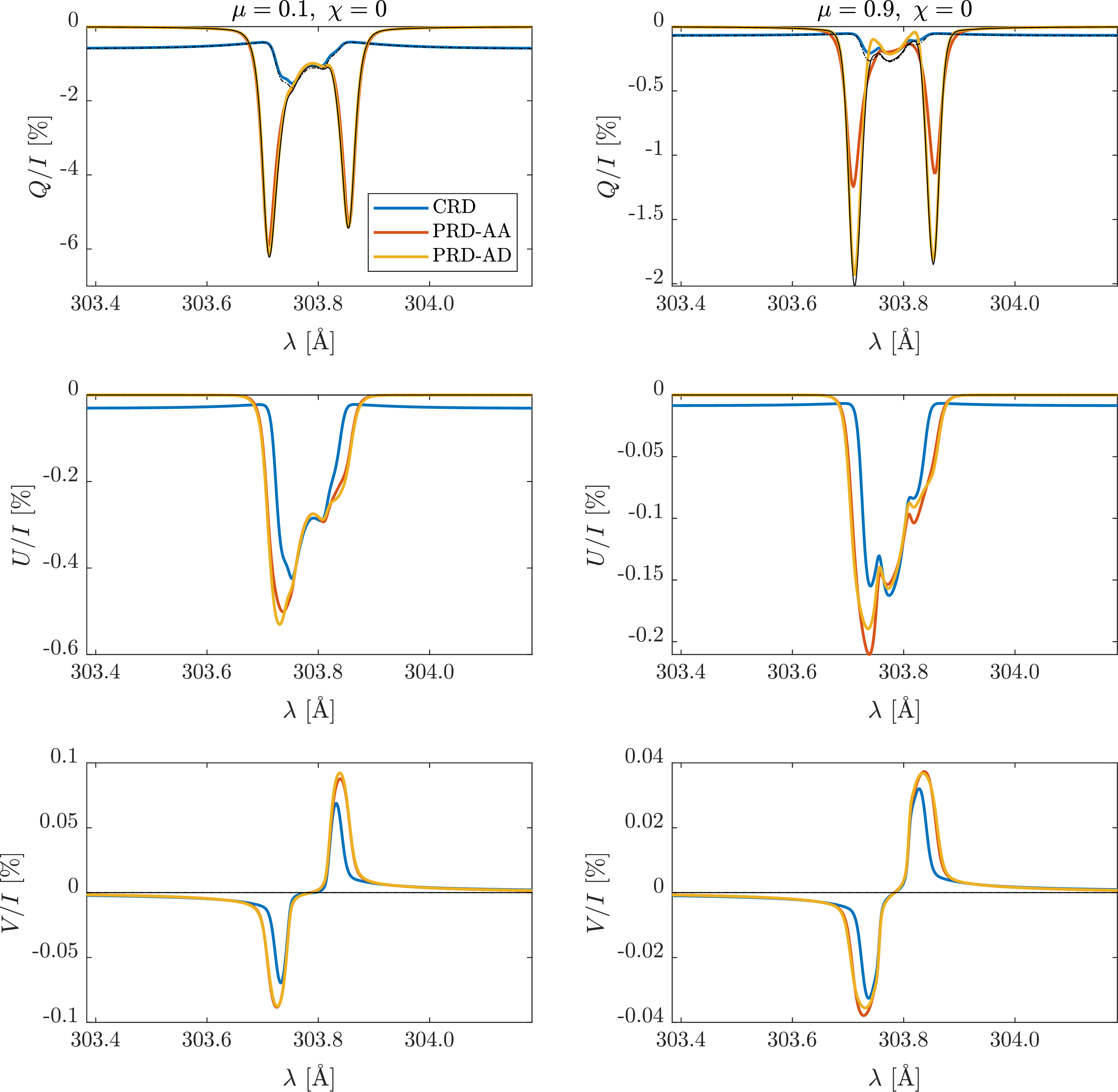}
    \caption{Stokes $Q/I$ (\emph{top panels}), $U/I$ (\emph{middle panels}), and $V/I$ (\emph{bottom panels}) profiles for the He~{\sc ii} Ly-$\alpha$ line as a function of wavelength. The results are obtained with the FAL-C atmospheric model, with a uniform horizontal ($\theta_B=\pi/2$ and $\chi_B=0$) magnetic field of 100 G, and for two LOSs, near the limb ($\mu=0.1$, \emph{left panels}) and near the disk centre ($\mu=0.9$, \emph{right panels}), respectively. Blue, red, and yellow lines correspond to 
    CRD, PRD--AA, and PRD--AD calculations, respectively. The unmagnetic CRD and PRD--AD cases are given for reference as black dash-dotted and solid lines, respectively.}
    \label{fig:QUV_B100}
\end{figure*}

The physical problem discussed in Sect.~\ref{sec:problem} and discretised as outlined in Sect.~\ref{sec:setting} was implemented as a \emph{MATLAB}~\citep{MATLAB:2023} routine and solved with the \emph{MATLAB} \emph{gmres} solver. 
In this section, we present the numerical results obtained
for the He~{\sc{ii}} Ly-$\alpha$ line with two main goals.
First, we aim at assessing the suitability of
the CRD and PRD--AA modellings through comparisons with the general PRD--AD calculations,
both in the absence and presence of magnetic fields.
Second, we investigate the potential of the 
wavelength-integrated scattering polarisation signal of this line
for probing the magnetism of the solar chromosphere.

\subsection{Impact of PRD--AD effects}
\label{sec:res_scattering_profiles}

The impact of the AD treatment on the emergent Stokes profiles 
is first analysed considering the FAL-C
atmospheric model and two lines of sight (LOSs), with $\mu=0.1$
and $\mu=0.9$. 
Figures~\ref{fig:I_QI_unmagnetic} and~\ref{fig:QUV_B100} expose
the synthetic emergent Stokes profiles in the absence and in the presence of a magnetic field, respectively. For the latter case, as an example of interest, we considered a uniform horizontal ($\theta_B=\pi/2$ and $\chi_B=0$) magnetic field of strength $B=100$~G.
In the unmagnetic case, the $U/I$ and $V/I$ profiles vanish and are consequently not shown. 
In addition, as the $I$ profile is not affected by magnetic fields of practical interest, it is only displayed for the unmagnetic case in Fig.~\ref{fig:I_QI_unmagnetic}.

In general,
we found a good agreement between the CRD, PRD--AA, and PRD--AD
calculations in the core of the line for all Stokes profiles and all magnetic conditions.
Reasonably, the CRD modelling is however not able to reproduce the
large peaks outside the line-core region of the $Q/I$ profiles. 
Noticeable discrepancies are also present near the disk centre ($\mu=0.9$) in the $Q/I$ peaks  between AA and AD calculations. 
This is not true near the limb ($\mu=0.1$), where we found an excellent agreement between AA and AD calculations. 
We also note that our $Q/I$ profiles are in 
good agreement with those shown in Fig.~3 of \citet{belluzzi2012b}.
This proves the robustness of our results, considering that the two works were carried out applying completely different solution 
strategies.\footnote{We recall that the results of \citet{belluzzi2012b} were obtained applying the redistribution matrix of \citet{belluzzi2014}. 
This differs from the one of \citet{bommier2017} in the $R^{\scriptscriptstyle \rm III}$ term, which was derived from heuristic arguments. 
This difference is, however, irrelevant in the modeling of He~{\sc ii} Ly-$\alpha$ because the contribution from $R^{\scriptscriptstyle \scriptscriptstyle \rm III}$ is completely 
negligible at the height where this line is formed \citep[see left panel of Fig. 1 of][]{belluzzi2012b}.}

Figure~\ref{fig:QUV_B100} also shows that, as expected, the $Q/I$ signal is barely sensitive
to the Hanle effect for the considered magnetic field
of 100~G (cf. the colour with the black lines).
However, a significant Hanle signal appears in the $U/I$ profile, particularly near the limb (see the central left panel).
Moreover, as expected because of the reduced effectiveness of the Zeeman effect at these wavelengths, the circular polarization signal is much weaker than the linear ones.
Notably, the amplitudes of the line-core $Q/I$ and $U/I$ signals are in excellent agreement with the CRD results shown in the left panel of Fig.~5 
of \citet{trujillo_bueno2012}.
In addition, the MO effects seem to play a marginal role for the considered cases.
In confirmation of this, we performed the same calculations
by forcing to zero the MO term of the
propagation matrix \citep[see][]{alsinaballester2017},
obtaining almost overlapping emergent $Q/I$ and $U/I$ signals. 
For the sake of completeness, we also carried out calculations analogous to Fig.~\ref{fig:QUV_B100}, but for vertical magnetic fields (not shown here for conciseness). For magnetic field strengths typical of the quiet solar chromosphere, the $Q/I$ signals did not display any noticeable variation with respect to the unmagnetic case, while the $U/I$ signals were extremely weak and therefore of reduced diagnostic interest.
 
A striking difference between the CRD and PRD results of Figs.~\ref{fig:I_QI_unmagnetic} and \ref{fig:QUV_B100} appears in the $I$, $Q/I$, and $U/I$ profiles far away from the line centre. 
In particular, the amplitude of the fractional polarisation signals at distances larger than 0.1~\AA~from the line centre is much larger for CRD than for PRD calculations.
A detailed investigation of these results, not reported here for conciseness, revealed that this different behaviour is not due to continuum contributions, which in this spectral region are extremely low. 
The difference is rather due to the strong redistribution effects resulting from the slowly decreasing wings of the Voigt emission profile entering the 
$R^{\scriptscriptstyle \rm III}$ redistribution matrix describing the CRD limit, combined to the very large and sharp emission peak of this line (moving away from the line centre,
the intensity decreases by more than four orders of magnitude in about 0.1\,{\AA}).\footnote{We repeated part of the above calculations considering a broader spectral domain $[\lambda_{\min},\lambda_{\max}]=[266.04\text{\,\AA},354.00\text{\,\AA}]$ and we discretised it with 381 spectral nodes. We verified that the continuum intensity and the $Q/I$ continuum polarisation obtained for the CRD and PRD treatments are the same, whereas $U/I$ goes to zero moving away from the line centre, irrespectively of the scattering description.} 
This leads to Stokes profiles that are much broader for the CRD than for the PRD case.

\subsection{Sensitivity of Stokes profiles to the atmospheric model}
\label{sec:res_atms}

\begin{figure*}[ht!]
    \centering
    \includegraphics[width=0.95\textwidth]{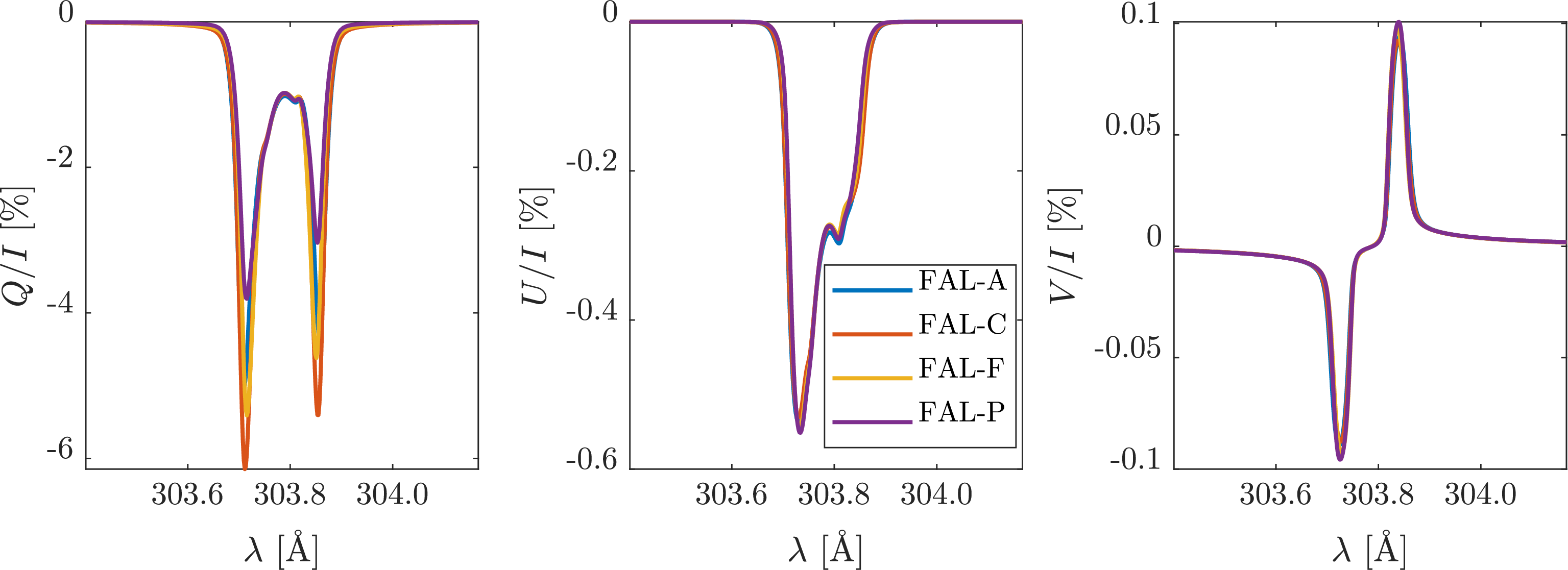}
    \caption{Stokes $Q/I$ (\emph{left panel}), $U/I$ (\emph{middle panel}), and $V/I$ (\emph{right panel}) profiles for the He~{\sc ii} Ly-$\alpha$ line as a function of wavelength. The results are obtained performing PRD-AD calculations for a LOS near the limb ($\mu=0.1$ and $\chi=0$) and with a uniform horizontal ($\theta_B=\pi/2$ and $\chi_B=0$) magnetic field of 100\,G. Blue, red, yellow, and purple lines correspond to profiles obtained with the FAL-A, FAL-C, FAL-F, and FAL-P atmospheric models, respectively.}
    \label{fig:QUV_FALACFP}
\end{figure*}

To analyse the sensitivity of the He~{\sc{ii}} Ly-$\alpha$
emergent Stokes profiles to atmospheric conditions, we considered
the four plane-parallel semi-empirical 1D FAL-A, FAL-C, FAL-F,
and FAL-P atmospheric models and we solved the RT problem
in the presence of a uniform horizontal ($\theta_B=\pi/2$ and $\chi_B=0$) magnetic field of 100~G.
Figure~\ref{fig:QUV_FALACFP} shows the emergent
$Q/I$, $U/I$, and $V/I$ profiles for the 
aforementioned atmospheric models and a LOS 
with $\mu=0.1$ and $\chi=0$.
We found a very good agreement in the line core of the $Q/I$ profile. On the other hand, some relevant differences appear in the peaks just outside the line-core region.
Moreover, the $U/I$ and $V/I$ signals almost overlap for all the considered atmospheric models.
We note that, although we only showed the $\mu=0.1$ case, good agreement was found for all LOSs. 
We also note that, although the present $Q/I$ profiles were obtained in a magnetised atmosphere, these are in excellent qualitative agreement with 
those in Fig. 5 by~\citet{belluzzi2012b}.

\subsection{Wavelength-integrated linear polarisation signals}
\label{sec:res:wintegarted}

\begin{figure*}[ht!]
    \centering
    \includegraphics[width=0.95\textwidth]{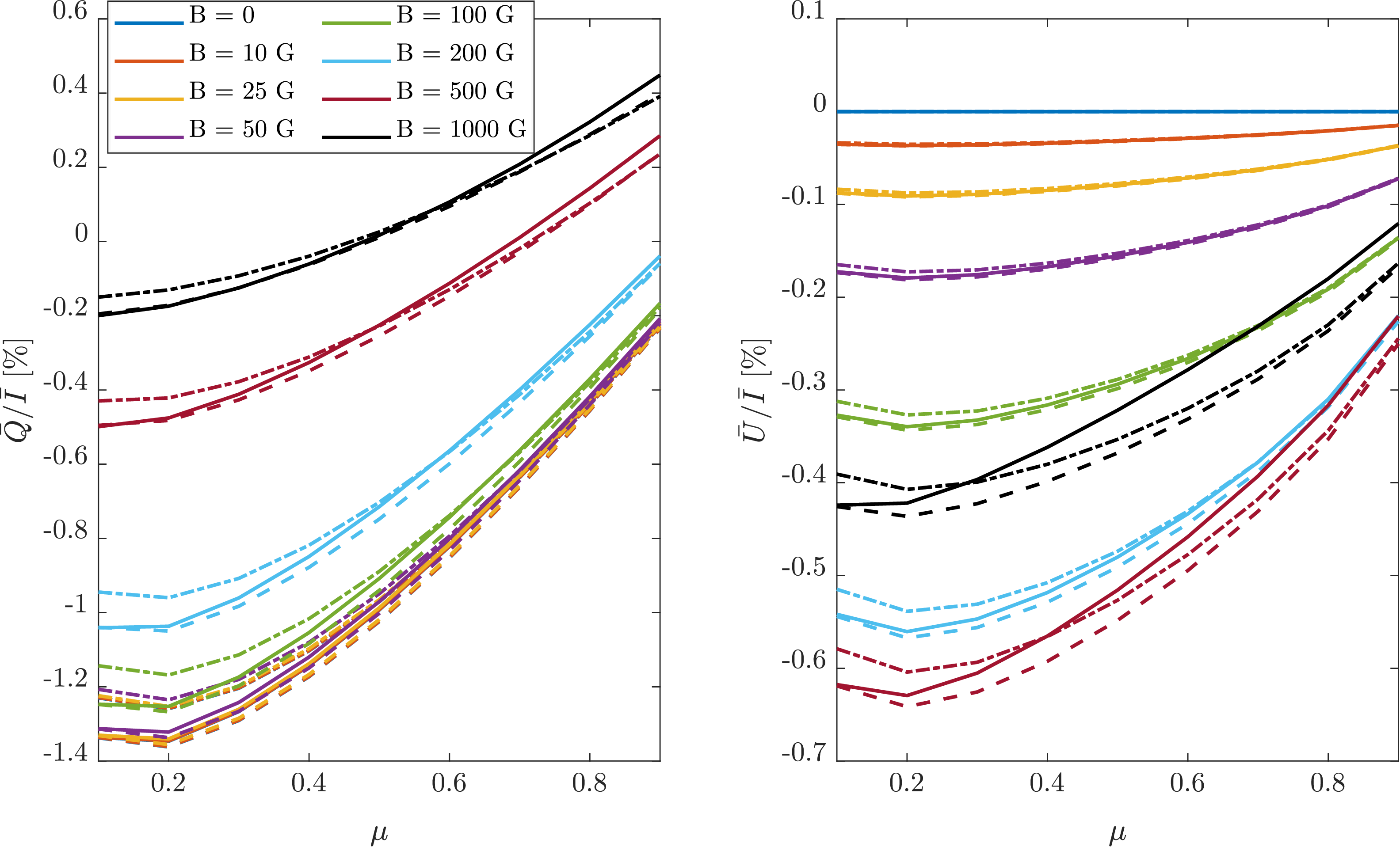}
    \caption{Centre-to-limb variation of $\overline{Q}/\overline{I}$ (\emph{left panel}) and $\overline{U}/\overline{I}$ (\emph{right panel}) signals as a function of $\mu=\cos(\theta)$. The results are obtained for $\chi=0$ and with the FAL-C atmospheric model, considering uniform horizontal ($\theta_B=\pi/2$ and $\chi_B=0$) magnetic fields of different strengths (see the legend). Dash-dotted, dashed, and solid lines correspond to CRD, PRD--AA, and PRD--AD 
    calculations, respectively.}
    \label{fig:wintegrated_diffB}
\end{figure*}
\begin{figure*}[ht!]
    \centering
    \includegraphics[width=0.95\textwidth]{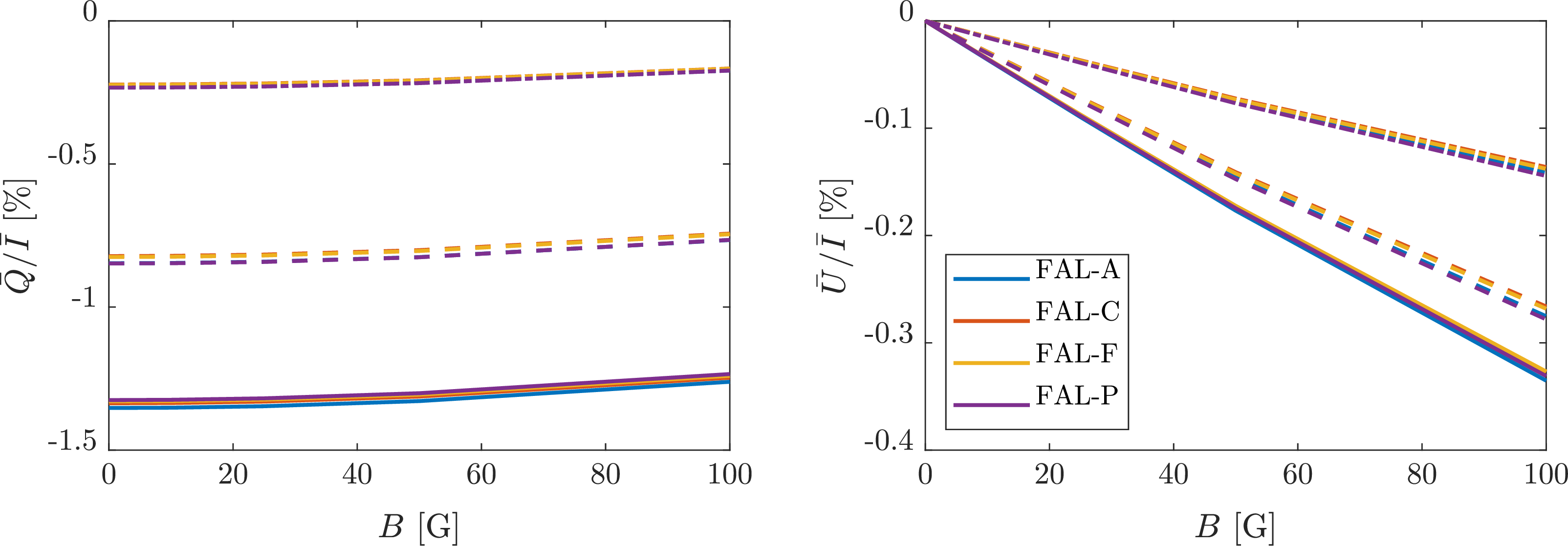}
    \caption{Plots of $\overline{Q}/\overline{I}$ (\emph{left panel}) and $\overline{U}/\overline{I}$ (\emph{right panel}) as a function of $B$. 
    The results are obtained performing PRD-AD calculations for three LOSs with $\mu=0.1,\,0.6,$ and 0.9 (solid, dashed, and dash-dotted lines, respectively), $\chi=0$, and considering uniform horizontal ($\theta_B=\pi/2$ and $\chi_B=0$) magnetic fields of different amplitudes. 
    Blue, red, yellow, and purple lines correspond to profiles obtained with the FAL-A, FAL-C, FAL-F, and FAL-P atmospheric model, respectively.}
    \label{fig:wintegrated_FALACFP}
\end{figure*}

We now focus on the He~{\sc{ii}} Ly-$\alpha$ linear polarisation signals obtained using narrowband filters.
Synthetic narrowband signals can be obtained by integrating over wavelength the numerical emergent Stokes profiles, weighted by a Gaussian that mimics the action of a narrowband filter. 
Namely, \citep{alsina_ballester2023}
\begin{equation}
\overline{X}(\mu)=\frac{1}{\sqrt{2\pi}\sigma_\lambda}
\int_{\lambda_0-\Delta\lambda}^{\lambda_0+\Delta\lambda}{\rm d}\lambda\exp\left(-\frac{(\lambda-\lambda_0)^2}{2\sigma_\lambda^2}\right)X(z_{\max},\mu,\lambda),
\end{equation}
where $X = I, Q, U$, $\lambda_0 = 303.784$~\AA~
is the He~{\sc{ii}} Ly-$\alpha$ line centre wavelength in vacuum,
and $\sigma_\lambda$
is the standard deviation of the
Gaussian weighting function, which
corresponds to a full width at half maximum $\text{FWHM} = 2\sqrt{2\log(2)}\sigma_\lambda$ (here we do not consider the circular polarization signal as its profile is symmetric and thus $\bar{V}$ vanishes).
We note that the choice of FWHM affects the
amplitude of the narrowband signals $\overline{I}$, $\overline{Q}$,
and $\overline{U}$, but it does not impact the ratios
$\overline{Q}/\overline{I}$ and $\overline{U}/\overline{I}$
and their dependency on $\mu$. In addition, any choice with $\Delta\lambda\geq0.5$ {\AA} has no impact on the results, because both the intensity and the scattering polarisation signals
quickly drop to zero moving away from the line centre. Thus, in the calculations, we adopted a FWHM of 1~\AA~and we used $\Delta\lambda=1$~\AA.

Figure~\ref{fig:wintegrated_diffB} displays
the CLV
of the narrowband $\overline{Q}/\overline{I}$ and $\overline{U}/\overline{I}$ signals for LOSs with $\mu\in[0.1,0.9]$,
considering the FAL-C atmospheric model
in the presence of uniform horizontal magnetic fields of strength between 0 and 1000~G.
For $B\leq200$~G, the amplitude of the $\overline{Q}/\overline{I}$ and $\overline{U}/\overline{I}$ signals increases from the disk centre to the limb for all the scattering descriptions, 
reaching a maximum close to $\mu=0.2$, and then slightly decreases towards $\mu=0.1$.
Moreover, the $\overline{Q}/\overline{I}$ signal
proves to be almost insensitive to magnetic fields of strengths $B\leq50$~G, while its amplitude decreases for $200~\mathrm{G}\geq B\geq50$~G due to the Hanle effect.
Conversely, the amplitude of the $\overline{U}/\overline{I}$ signal increases almost linearly with $B$
for all the considered inclinations.
For $B>200$~G, the amplitude of $\overline{U}/\overline{I}$ reaches a maximum near $B=500$~G, and then decreases for stronger magnetic fields due to Hanle rotation.
Furthermore, we note that $\bar{Q}/\bar{I}$ signals become positive approaching the disk centre, with a larger amplitude for stronger magnetic fields. This is due to the forward-scattering Hanle effect~\citep{trujillo_bueno2002}.
In general, we found a very good agreement between CRD, PRD--AA, and PRD--AD calculations, although
the CRD description slightly underestimates the amplitude of the PRD results, in particular for $\mu<0.5$.
Noticeably, the agreement between PRD--AA and PRD--AD calculations decreases near the disk centre for $B\geq500$~G.

To investigate the dependence of the synthetic narrowband signals on atmospheric parameters, in Fig.~\ref{fig:wintegrated_FALACFP} we present
$\overline{Q}/\overline{I}$ and $\overline{U}/\overline{I}$ as a function of
the strength of the horizontal magnetic field
for the three different LOSs $\mu=0.1,\,0.6,$ and 0.9
and for the four aforementioned FAL atmospheric models.
The dependence of all signals on the considered atmospheric models is negligible.
Moreover, the left panel of Fig.~\ref{fig:wintegrated_FALACFP} shows that $\overline{Q}/\overline{I}$ is almost flat for $B<50$ G, 
while it reveals a slight Hanle depolarisation for stronger magnetic fields, in agreement with the results presented in Fig.~\ref{fig:wintegrated_diffB}.
In addition, in the right panel of Fig.~\ref{fig:wintegrated_FALACFP} we see that the amplitude of the appreciable $\overline{U}/\overline{I}$ signal
increases almost linearly with $B$, also in agreement with the results in Fig.~\ref{fig:wintegrated_diffB}.

\section{Discussion and conclusions}
\label{sec:conclusions}
In Sect.~\ref{sec:res:wintegarted},
we showed that synthetic narrowband signals of the He~{\sc{ii}} Ly-$\alpha$ line are insensitive to the considered atmospheric models. At the same time, a study of the impact of bulk velocities on the He~{\sc{ii}} Ly-$\alpha$ line, not reported here for conciseness, revealed that $\overline{Q}/\overline{I}$ and $\overline{U}/\overline{I}$ are also insensitive to
typical chromospheric plasma bulk velocities.\footnote{We repeated part of the calculations presented above for a vertical bulk velocity with a uniform gradient of 0.009 1/s. The results do not display any noticeable difference.} 
We also showed that, although the Stokes profiles are nearly immune to MO effects, $\overline{U}/\overline{I}$ displays an almost linear dependence on $B$, via the Hanle effect, for magnetic strengths typical of the solar chromosphere, and that $\overline{Q}/\overline{I}$ is almost insensitive to 
horizontal magnetic fields of strength weaker than 50\,G.
Consequently, since $|\overline{U}/\overline{I}|>0.3\%$ near the solar limb for a horizontal magnetic field of 100~G, He~{\sc{ii}} Ly-$\alpha$ linear polarisation signals obtained using narrowband filters may represent an interesting novel capability for inferring chromospheric magnetic fields, provided that a sufficient signal-to-noise ratio can be achieved.

It is thus important to assess what is the integration time, $t_\text{int}$, necessary to be sensitive to chromospheric magnetic fields for a given instrument. 
In this respect, we applied Eq.~(56) in~\citet{trujillo_bueno2017} considering: a near-limb LOS, for which $|\overline{U}/\overline{I}|$ and $|\overline{Q}/\overline{I}|$ are maximal; a polarimetric sensitivity of $0.05\,\%$ that, in the present geometry, means a sensitivity to horizontal magnetic fields of about 10~G; an instrumental efficiency of $1\,\%$; a spatial resolution of 2 arcsec; a diameter of the telescope of 30 cm, similar to the instrument installed on the CLASP experiment~\citep{kobayashi2012}; and a FWHM of the narrowband filter of 1\,{\AA}.
Given these constraints, we found that the integration time necessary to be sensitive to horizontal chromospheric magnetic fields of 10~G is $t_\text{int}=885$, 447, 157, and 22 s for the atmospheric models FAL-A, FAL-C, FAL-F, and FAL-P, respectively. We note that, while $\overline{Q}/\overline{I}$ and $\overline{U}/\overline{I}$ are independent of the atmospheric model, $t_\text{int}$ is not, as the intensity signal, and thus the number of photons reaching the telescope, strongly depend on the atmospheric temperature structure. 

Concerning quite-solar conditions (i.e. models FAL-A, FAL-C, and FAL-F), the resulting $t_\text{int}$ are too large to be reached through 
sounding rocket experiments. 
On the other hand, the integration time for the FAL-P model, which corresponds to a typical plage area, is realistic for present days' rockets (e.g. CLASP).
Overall, valuable He~{\sc{ii}} Ly-$\alpha$ polarization signals could also be measured in quiet regions, across 
the whole solar disk, with spacecraft missions, as these allow for considerably larger telescopes and longer exposure times than sounding rocket experiments.

Another interesting outcome of the work is the very good agreement of wavelength integrated He~{\sc{ii}} Ly-$\alpha$ fractional linear polarisation signals obtained from CRD, PRD--AA, and PRD--AD calculations, especially for magnetic fields up to 200 G. 
This means that a complete PRD--AD description of scattering processes seems not to be strictly necessary to accurately model the impact of such magnetic fields on these signals. 
The CRD formulation could be enough. This would drastically reduce the computational cost of inferring magnetic fields from future He~{\sc{ii}} Ly-$\alpha$ observations.

We remind that the results in the present paper were obtained using a semi-empirical 1D model. 
However, the solar atmosphere is definitely not 1D, but it has large horizontal inhomogeneities as well as bulk velocities with strong gradients, 
which can strongly impact the amplitude and shape of scattering polarisation signals \citep[e.g.][]{mansosainz2011,stepan2016,jaumebestard2021}.
This is particularly true in the upper chromosphere. 
Consequently, in the future it will be essential to assess the impact of these effects on 
the polarisation signals of the He~{\sc{ii}} Ly-$\alpha$ line through full 3D RT calculations in state-of-the-art models of the solar atmosphere.
The same 3D calculations, to be performed both in CRD \citep[e.g. using the PORTA code of][]{stepan2013} and in PRD \citep[e.g. using a code based on the solution strategy outlined in][]{benedusi2023}, would confirm whether also in the presence of such effects, the limit of CRD remains a good approximation for modeling the 
narrowband signals of this line.

This work highlights the diagnostic potential of filter-polarimetry in the He~{\sc{ii}} Ly-$\alpha$ line.
Our 1D results, to be confirmed by more accurate 3D calculations, suggest that in near-limb plage regions the sensitivity of
narrowband He~{\sc{ii}} Ly-$\alpha$ linear polarisation signals to chromospheric magnetic fields can already be assessed through sounding rocket experiments.
Given its minor sensitivity to horizontal magnetic fields, $\overline{Q}/\overline{I}$ could be used as a reference, while $\overline{U}/\overline{I}$ would be 
exploited to infer information on the magnetic field in the upper chromospheric layers.
Notably, given the very narrow region in which the He~{\sc{ii}} Ly-$\alpha$ line forms~\citep{trujillo_bueno2012,belluzzi2012b}, we would also have strong constraints 
on the precise location of the estimated magnetic fields.


\begin{acknowledgements}
The financial support by the Swiss National Science Foundation (SNSF) through grant CRSII5\_180238 is gratefully acknowledged.
\end{acknowledgements}

\bibliographystyle{aa}
\bibliography{bibfile}

\begin{thebibliography}{45}
\expandafter\ifx\csname natexlab\endcsname\relax\def\natexlab#1{#1}\fi

\bibitem[{{Alsina Ballester} {et~al.}(2016){Alsina Ballester}, {Belluzzi}, \&
  {Trujillo Bueno}}]{alsinaballester2016}
{Alsina Ballester}, E., {Belluzzi}, L., \& {Trujillo Bueno}, J. 2016, \apjl,
  831, L15

\bibitem[{{Alsina Ballester} {et~al.}(2017){Alsina Ballester}, {Belluzzi}, \&
  {Trujillo Bueno}}]{alsinaballester2017}
{Alsina Ballester}, E., {Belluzzi}, L., \& {Trujillo Bueno}, J. 2017, \apj,
  836, 6

\bibitem[{{Alsina Ballester} {et~al.}(2018){Alsina Ballester}, {Belluzzi}, \&
  {Trujillo Bueno}}]{alsinaballester2018}
{Alsina Ballester}, E., {Belluzzi}, L., \& {Trujillo Bueno}, J. 2018, \apj,
  854, 150

\bibitem[{{Alsina Ballester} {et~al.}(2021){Alsina Ballester}, {Belluzzi}, \&
  {Trujillo Bueno}}]{alsinaballester2021}
{Alsina Ballester}, E., {Belluzzi}, L., \& {Trujillo Bueno}, J. 2021, \prl,
  127, 081101

\bibitem[{{Alsina Ballester} {et~al.}(2022){Alsina Ballester}, {Belluzzi}, \&
  {Trujillo Bueno}}]{alsinaballester2022}
{Alsina Ballester}, E., {Belluzzi}, L., \& {Trujillo Bueno}, J. 2022, \aap,
  664, A76

\bibitem[{{Alsina Ballester} {et~al.}(2023){Alsina Ballester}, {Belluzzi}, \&
  {Trujillo Bueno}}]{alsina_ballester2023}
{Alsina Ballester}, E., {Belluzzi}, L., \& {Trujillo Bueno}, J. 2023, \apj,
  947, 71

\bibitem[{{Belluzzi} \& {Trujillo Bueno}(2014)}]{belluzzi2014}
{Belluzzi}, L. \& {Trujillo Bueno}, J. 2014, \aap, 564, A16

\bibitem[{{Belluzzi} {et~al.}(2012){Belluzzi}, {Trujillo Bueno}, \&
  {{\v{S}}t{\v{e}}p{\'a}n}}]{belluzzi2012b}
{Belluzzi}, L., {Trujillo Bueno}, J., \& {{\v{S}}t{\v{e}}p{\'a}n}, J. 2012,
  \apjl, 755, L2

\bibitem[{{Benedusi} {et~al.}(2021){Benedusi}, {Janett}, {Belluzzi}, \&
  {Krause}}]{benedusi2021}
{Benedusi}, P., {Janett}, G., {Belluzzi}, L., \& {Krause}, R. 2021, \aap, 655,
  A88

\bibitem[{{Benedusi} {et~al.}(2022){Benedusi}, {Janett}, {Riva}, {Belluzzi}, \&
  {Krause}}]{benedusi2022}
{Benedusi}, P., {Janett}, G., {Riva}, G., {Belluzzi}, L., \& {Krause}, R. 2022,
  \aap, 664, A197

\bibitem[{{Benedusi} {et~al.}(2023){Benedusi}, {Riva}, {Zulian},
  {{\v{S}}t{\v{e}}p{\'a}n}, {Belluzzi}, \& {Krause}}]{benedusi2023}
{Benedusi}, P., {Riva}, S., {Zulian}, P., {et~al.} 2023, Journal of
  Computational Physics, 479, 112013

\bibitem[{{Bommier}(1997)}]{bommier1997b}
{Bommier}, V. 1997, \aap, 328, 726

\bibitem[{{Bommier}(2017)}]{bommier2017}
{Bommier}, V. 2017, \aap, 607, A50

\bibitem[{{del Pino Alem{\'a}n} {et~al.}(2016){del Pino Alem{\'a}n}, {Casini},
  \& {Manso Sainz}}]{delpinoaleman2016}
{del Pino Alem{\'a}n}, T., {Casini}, R., \& {Manso Sainz}, R. 2016, \apjl, 830,
  L24

\bibitem[{{Fontenla} {et~al.}(1991){Fontenla}, {Avrett}, \&
  {Loeser}}]{fontenla1991}
{Fontenla}, J.~M., {Avrett}, E.~H., \& {Loeser}, R. 1991, \apj, 377, 712

\bibitem[{{Fontenla} {et~al.}(1993){Fontenla}, {Avrett}, \&
  {Loeser}}]{fontenla1993}
{Fontenla}, J.~M., {Avrett}, E.~H., \& {Loeser}, R. 1993, \apj, 406, 319

\bibitem[{{Janett} {et~al.}(2021){Janett}, {Ballester}, {Guerreiro}, {Riva},
  {Belluzzi}, {del Pino Alem{\'a}n}, \& {Bueno}}]{janett2021b}
{Janett}, G., {Ballester}, E.~A., {Guerreiro}, N., {et~al.} 2021, \aap, 655,
  A13

\bibitem[{Janett {et~al.}(2023)Janett, Benedusi, \& Riva}]{janett2023}
Janett, G., Benedusi, P., \& Riva, F. 2023, Astronomy {\&} Astrophysics

\bibitem[{Janett {et~al.}(2017)Janett, Carlin, Steiner, \&
  Belluzzi}]{janett2017formal}
Janett, G., Carlin, E.~S., Steiner, O., \& Belluzzi, L. 2017, The Astrophysical
  Journal, 840, 107

\bibitem[{Janett \& Paganini(2018)}]{janett2018formal}
Janett, G. \& Paganini, A. 2018, The Astrophysical Journal, 857, 91

\bibitem[{{Janett} {et~al.}(2018){Janett}, {Steiner}, \&
  {Belluzzi}}]{janett2018formal2}
{Janett}, G., {Steiner}, O., \& {Belluzzi}, L. 2018, The Astrophysical Journal,
  865, 16

\bibitem[{{Janev} {et~al.}(1987){Janev}, {Langer}, \& {Evans}}]{janev1987}
{Janev}, R.~K., {Langer}, W.~D., \& {Evans}, K. 1987, {Elementary processes in
  Hydrogen-Helium plasmas - Cross sections and reaction rate coefficients}
  (Springer)

\bibitem[{{Jaume Bestard} {et~al.}(2021){Jaume Bestard}, {Trujillo Bueno},
  {{\v{S}}t{\v{e}}p{\'a}n}, \& {del Pino Alem{\'a}n}}]{jaumebestard2021}
{Jaume Bestard}, J., {Trujillo Bueno}, J., {{\v{S}}t{\v{e}}p{\'a}n}, J., \&
  {del Pino Alem{\'a}n}, T. 2021, \apj, 909, 183

\bibitem[{{Kano} {et~al.}(2012){Kano}, {Bando}, {Narukage}, {Ishikawa},
  {Tsuneta}, {Katsukawa}, {Kubo}, {Ishikawa}, {Hara}, {Shimizu}, {Suematsu},
  {Ichimoto}, {Sakao}, {Goto}, {Kato}, {Imada}, {Kobayashi}, {Holloway},
  {Winebarger}, {Cirtain}, {De Pontieu}, {Casini}, {Trujillo Bueno},
  {{\v{S}}tep{\'a}n}, {Manso Sainz}, {Belluzzi}, {Asensio Ramos},
  {Auch{\`e}re}, \& {Carlsson}}]{kano2012}
{Kano}, R., {Bando}, T., {Narukage}, N., {et~al.} 2012, in Society of
  Photo-Optical Instrumentation Engineers (SPIE) Conference Series, Vol. 8443,
  Space Telescopes and Instrumentation 2012: Ultraviolet to Gamma Ray, ed.
  T.~{Takahashi}, S.~S. {Murray}, \& J.-W.~A. {den Herder}, 84434F

\bibitem[{{Kano} {et~al.}(2017){Kano}, {Trujillo Bueno}, {Winebarger},
  {Auch{\`e}re}, {Narukage}, {Ishikawa}, {Kobayashi}, {Bando}, {Katsukawa},
  {Kubo}, {Ishikawa}, {Giono}, {Hara}, {Suematsu}, {Shimizu}, {Sakao},
  {Tsuneta}, {Ichimoto}, {Goto}, {Belluzzi}, {{\v{S}}t{\v{e}}p{\'a}n}, {Asensio
  Ramos}, {Manso Sainz}, {Champey}, {Cirtain}, {De Pontieu}, {Casini}, \&
  {Carlsson}}]{kano2017}
{Kano}, R., {Trujillo Bueno}, J., {Winebarger}, A., {et~al.} 2017, \apjl, 839,
  L10

\bibitem[{{Kobayashi} {et~al.}(2012){Kobayashi}, {Kano}, {Trujillo-Bueno},
  {Asensio Ramos}, {Bando}, {Belluzzi}, {Carlsson}, {De Pontieu}, {Hara},
  {Ichimoto}, {Ishikawa}, {Katsukawa}, {Kubo}, {Manso Sainz}, {Narukage},
  {Sakao}, {Stepan}, {Suematsu}, {Tsuneta}, {Watanabe}, \&
  {Winebarger}}]{kobayashi2012}
{Kobayashi}, K., {Kano}, R., {Trujillo-Bueno}, J., {et~al.} 2012, in
  Astronomical Society of the Pacific Conference Series, Vol. 456, Fifth Hinode
  Science Meeting, ed. L.~{Golub}, I.~{De Moortel}, \& T.~{Shimizu}, 233

\bibitem[{{Landi Degl'Innocenti} \&
  {Landolfi}(2004)}]{landi_deglinnocenti+landolfi2004}
{Landi Degl'Innocenti}, E. \& {Landolfi}, M. 2004, Astrophysics and Space
  Science Library, Vol. 307, {Polarization in Spectral Lines} (Dordrecht:
  Kluwer Academic Publishers)

\bibitem[{{Leenaarts} {et~al.}(2012){Leenaarts}, {Pereira}, \&
  {Uitenbroek}}]{leenaarts2012}
{Leenaarts}, J., {Pereira}, T., \& {Uitenbroek}, H. 2012, \aap, 543, A109

\bibitem[{{Manso Sainz} \& {Trujillo Bueno}(2011)}]{mansosainz2011}
{Manso Sainz}, R. \& {Trujillo Bueno}, J. 2011, \apj, 743, 12

\bibitem[{MATLAB(2023)}]{MATLAB:2023}
MATLAB. 2023, version 9.14.0 (R2023a) (Natick, Massachusetts: The MathWorks
  Inc.)

\bibitem[{Mihalas(1978)}]{mihalas1978}
Mihalas, D. 1978, Stellar Atmospheres, 2nd edn. (San Francisco: W.H.~Freeman
  and Company)

\bibitem[{{Riva} {et~al.}(2023){Riva}, {Guerreiro}, {Janett}, {Rossinelli},
  {Benedusi}, {Krause}, \& {Belluzzi}}]{riva2023}
{Riva}, S., {Guerreiro}, N., {Janett}, G., {et~al.} 2023, \aap, 679, A87

\bibitem[{{Sampoorna} {et~al.}(2017){Sampoorna}, {Nagendra}, \&
  {Stenflo}}]{sampoorna2017}
{Sampoorna}, M., {Nagendra}, K.~N., \& {Stenflo}, J.~O. 2017, \apj, 844, 97

\bibitem[{{Trujillo Bueno}(2003)}]{trujillo_bueno2003}
{Trujillo Bueno}, J. 2003, in Astronomical Society of the Pacific Conference
  Series, Vol. 288, Stellar Atmosphere Modeling, ed. I.~{Hubeny}, D.~{Mihalas},
  \& K.~{Werner}, 551

\bibitem[{{Trujillo Bueno} \& {del Pino Alem{\'a}n}(2022)}]{trujillo_bueno2022}
{Trujillo Bueno}, J. \& {del Pino Alem{\'a}n}, T. 2022, \araa, 60, 415

\bibitem[{{Trujillo Bueno} {et~al.}(2017){Trujillo Bueno}, {Landi
  Degl'Innocenti}, \& {Belluzzi}}]{trujillo_bueno2017}
{Trujillo Bueno}, J., {Landi Degl'Innocenti}, E., \& {Belluzzi}, L. 2017, \ssr,
  210, 183

\bibitem[{{Trujillo Bueno} {et~al.}(2002){Trujillo Bueno}, {Landi
  Degl'Innocenti}, {Collados}, {Merenda}, \& {Manso
  Sainz}}]{trujillo_bueno2002}
{Trujillo Bueno}, J., {Landi Degl'Innocenti}, E., {Collados}, M., {Merenda},
  L., \& {Manso Sainz}, R. 2002, \nat, 415, 403

\bibitem[{{Trujillo Bueno} {et~al.}(2012){Trujillo Bueno},
  {{\v{S}}t{\v{e}}p{\'a}n}, \& {Belluzzi}}]{trujillo_bueno2012}
{Trujillo Bueno}, J., {{\v{S}}t{\v{e}}p{\'a}n}, J., \& {Belluzzi}, L. 2012,
  \apjl, 746, L9

\bibitem[{{Trujillo Bueno} {et~al.}(2018){Trujillo Bueno},
  {{\v{S}}t{\v{e}}p{\'a}n}, {Belluzzi}, {Asensio Ramos}, {Manso Sainz}, {del
  Pino Alem{\'a}n}, {Casini}, {Ishikawa}, {Kano}, {Winebarger}, {Auch{\`e}re},
  {Narukage}, {Kobayashi}, {Bando}, {Katsukawa}, {Kubo}, {Ishikawa}, {Giono},
  {Hara}, {Suematsu}, {Shimizu}, {Sakao}, {Tsuneta}, {Ichimoto}, {Cirtain},
  {Champey}, {De Pontieu}, \& {Carlsson}}]{trujillo_bueno2018}
{Trujillo Bueno}, J., {{\v{S}}t{\v{e}}p{\'a}n}, J., {Belluzzi}, L., {et~al.}
  2018, \apjl, 866, L15

\bibitem[{{Trujillo Bueno} {et~al.}(2011){Trujillo Bueno},
  {{\v{S}}t{\v{e}}p{\'a}n}, \& {Casini}}]{trujillo_bueno2011}
{Trujillo Bueno}, J., {{\v{S}}t{\v{e}}p{\'a}n}, J., \& {Casini}, R. 2011,
  \apjl, 738, L11

\bibitem[{{Uitenbroek}(2001)}]{uitenbroek2001}
{Uitenbroek}, H. 2001, \apj, 557, 389

\bibitem[{{{\v S}t{\v e}p{\'a}n} \& {Trujillo Bueno}(2013)}]{stepan2013}
{{\v S}t{\v e}p{\'a}n}, J. \& {Trujillo Bueno}, J. 2013, \aap, 557, A143

\bibitem[{{{\v{S}}t{\v{e}}p{\'a}n} \& {Trujillo Bueno}(2016)}]{stepan2016}
{{\v{S}}t{\v{e}}p{\'a}n}, J. \& {Trujillo Bueno}, J. 2016, \apjl, 826, L10

\bibitem[{{{\v{S}}t{\v{e}}p{\'a}n} {et~al.}(2012){{\v{S}}t{\v{e}}p{\'a}n},
  {Trujillo Bueno}, {Carlsson}, \& {Leenaarts}}]{stepan2012}
{{\v{S}}t{\v{e}}p{\'a}n}, J., {Trujillo Bueno}, J., {Carlsson}, M., \&
  {Leenaarts}, J. 2012, \apjl, 758, L43

\bibitem[{{{\v{S}}t{\v{e}}p{\'a}n} {et~al.}(2015){{\v{S}}t{\v{e}}p{\'a}n},
  {Trujillo Bueno}, {Leenaarts}, \& {Carlsson}}]{stepan2015}
{{\v{S}}t{\v{e}}p{\'a}n}, J., {Trujillo Bueno}, J., {Leenaarts}, J., \&
  {Carlsson}, M. 2015, \apj, 803, 65

\end{thebibliography}


\end{document}